\documentclass[aps,prl,reprint,superscriptaddress]{revtex4-2}
\usepackage[utf8]{inputenc}
\usepackage{graphicx}
\usepackage[english]{babel}
\usepackage[T1]{fontenc}
\usepackage{amssymb}
\usepackage{amsmath}
\usepackage{caption}
\usepackage{subcaption}
\usepackage{comment}

\usepackage{color}
\usepackage{xcolor}

\usepackage{hyperref}
\usepackage{url}
\hypersetup{
    colorlinks,
    linkcolor={red!50!black},
    citecolor={blue!50!black},
    urlcolor={blue!80!black}
  }

\begin{document}


\title{Universal wrinkling of supported elastic rings}

\author{Benjamin Foster}
\email{ben\_foster@berkeley.edu}
\affiliation{Department of Physics, University of California at Berkeley, Berkeley,
  California 94720, USA}

\author{Nicol\'{a}s Verschueren}
\email{nverschueren@berkeley.edu}
\affiliation{Department of Physics, University of California at Berkeley, Berkeley,
California 94720, USA}
\affiliation{College of Engineering, Mathematics and Physical Sciences, University of Exeter, Exeter, United Kingdom}

\author{Edgar Knobloch}
\email{knobloch@berkeley.edu}
\affiliation{Department of Physics, University of California at Berkeley, Berkeley,
California 94720, USA}

\author{Leonardo Gordillo}
\email{leonardo.gordillo@usach.cl}
\affiliation{Departamento de F\'{i}sica, Facultad de Ciencia, Universidad de Santiago
de Chile, Chile}

\begin{abstract}
An exactly solvable family of models describing the wrinkling of substrate-supported inextensible elastic rings under compression is identified. The resulting wrinkle profiles are shown to be related to the buckled states of an unsupported ring and are therefore universal. Closed analytical expressions for the resulting universal shapes are provided, including the one-to-one relations between the pressure and tension at which these emerge. The analytical predictions agree with numerical continuation results to within numerical accuracy, for a large range of parameter values, up to the point of self-contact.
\end{abstract}

\maketitle

In-plane buckling of inextensible elastic rings under pressure has been studied over many years \cite{levy1884,tadjbakhsh_equilibrium_1967,flaherty_1972,flaherty_1973} and exact expressions for the buckled profiles are known \cite{Arreaga_shapes,Djondjorov_equil_shapes,djondjorov_ICGIQ}. Recent interest has centered on the effects of a supporting substrate. The inclusion of substrate forces leads to the emergence of an intrinsic scale $\lambda$. When compressed, a substrate-supported ring wrinkles with a critical wavelength defined by this scale instead of simply buckling \cite{diamant_witten2011,pocivavsek2008,Brau:2013jn,rivetti2013,veerapaneni2015,LeoEdgarsheet}. Periodic buckled and wrinkled states may emerge quasistatically, for example, in externally confined rings \cite{hazel_mullin_2017} or crumpled spherical shells \cite{vliegenthart_compression_2011}, in centrifugally or magnetically driven interfacial fingering in a Hele-Shaw cell \cite{Carvalho2013_HS_interfacial,Carvalho2014_HS_elasticA,Carvalho2014_HS_elasticB,Livera_magnetoelastic,Carrillo1996} and in the swelling of water-lecithin vesicles \cite{harbich_optical_1977,harbich_swelling_1984}, but may also arise dynamically, for example, during the dynamic collapse of an elastic ring around a soap film \cite{box2020}, the dynamic wrinkling of compressed floating elastica \cite{box2019} or in pulsating blood vessels \cite{VEERAPANENI2009}. The spatial profiles present in these very different systems are often strikingly similar, and this similarity remains unexplored.


Recent work on a family of simple, yet realistic, models for substrate-supported elastic rings under compression \cite{foster_pressure-driven_2022} revealed that these models have a special structure that suggests that {\it exact} wrinkle solutions can be constructed, and that these may, in turn, be related to the well-known buckled states of the substrate-free case.
In this Letter we show that this is indeed the case. Specifically, we show that, for this family of substrate forces, the wrinkle profiles generated by compression are related to the buckled states of the free, unsupported ring \cite{Arreaga_shapes}. We thereby show that the resulting wrinkle profiles are {\it universal} for this set of substrate forces. We determine the parameter space mapping that relates the buckled solutions of the classical, unsupported ring problem to the wrinkle solutions for rings with substrate support.  We use this mapping to predict bifurcation diagrams for this class of supported-ring problems and test the predictions via numerical continuation.

\begin{figure}[h]
    \centering
    \includegraphics[width=0.85\linewidth]{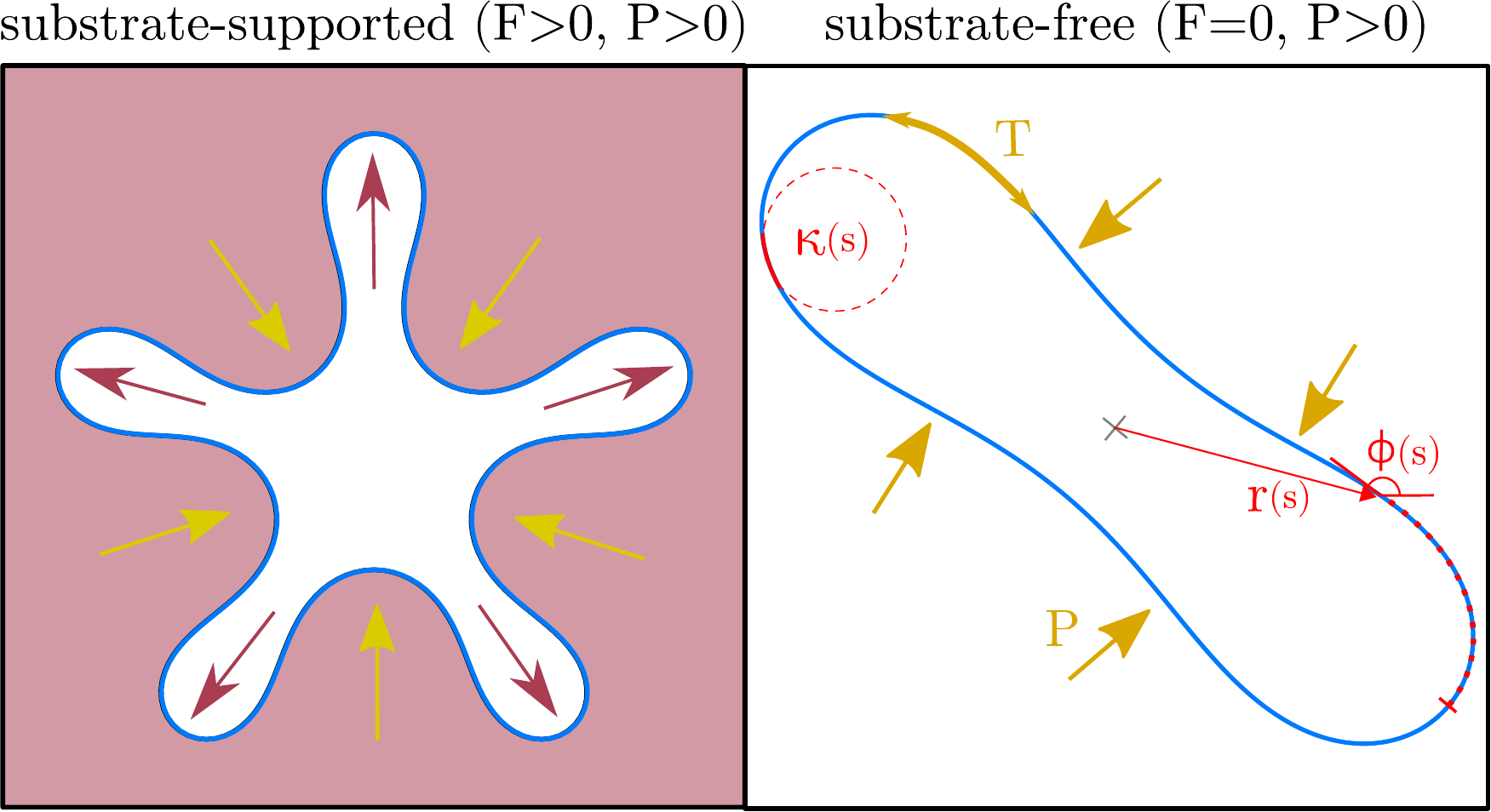}
    \caption{Possible regimes of interest. The blue curves represent the solution profiles $\mathbf{r}(s)$. Left panel: competition between pressure and substrate forces ($F>0$) leads to a nontrivial critical wrinkle wave number $m=5$. Right panel: the substrate-free ($F=0$) case leads to a buckled state with $m=2$. The problem variables are shown in the right panel.}
    \label{fig:substra}
\end{figure}

In order to study the wrinkling of a thin elastic inextensible ring supported by a soft substrate, we consider the following model, which can be derived from the Kirchhoff equations for elastic rods \cite{foster_pressure-driven_2022,audoly_pomeau}:
\begin{align}
       & \partial_s^2 \kappa + \frac{1}{2}\kappa^3 - T \kappa - P - \frac{1}{2} F(r) = 0, \nonumber \\
       &\quad\kappa(s)\equiv \partial_s \phi, \quad \partial_s{\bf r} = (\cos \phi, \sin \phi).
       \label{eq:phieq}
\end{align}
As shown in Fig.~\ref{fig:substra}, $\phi(s)$ is the local angle relative to the $x$ axis, $s$ is the arclength along the
solution profile with length $2\pi R$, $\kappa(s)$ is the curvature, and ${\bf r}(s)\equiv(x(s),y(s))$ is the radial distance from the ring center
to the point $s$. The boundary conditions $\phi(2\pi R)=\phi(0)+2\pi$ and the continuity of $x$, $y$,
$\partial_s \phi$ at $s=0$ and $2\pi R$ rule out non-smooth solutions. The quantities $P$ and $F(r)$ are the pressure
load inwards across the ring and the external force per unit of surface due to the substrate, respectively
(Fig.~\ref{fig:substra}). The Lagrange multiplier $T$ imposes inextensibility and is a nonlinear eigenvalue related
to the tension $\tau$ by $T=\tau+\frac{3}{2}\kappa^2$. In particular, we study exact solutions for
$F_n(r)=\alpha_n (r^n-r_0^n),\, n=0,2,4,6$, where the constant term $\alpha_n r_0^n/2$ is absorbed into $P$. Our solutions
also describe rings tethered to ${\bf r}=0$, provided $F_n(r)=-\alpha_nr^n<0$ and $P<0$ (interior overpressure).

In recent work \cite{foster_pressure-driven_2022}, the
case $n=2$ was taken as a simple model able to
capture the wrinkle-to-smooth transition that takes place in the
endothelium of an artery as the internal blood pressure oscillates. In
this case $\alpha_2=({R}/\lambda)^5$, where $\lambda\equiv ({\cal B}/K)^{1/5}$
is the bending length scale. Here ${\cal B}$ is the bending
modulus of the endothelium lining and $K$ is the arterial substrate
stiffness. The equations also describe the wrinkling of a
circular elastic membrane separating a higher density interior fluid
from a lower density exterior fluid in a rotating
Hele-Shaw cell
\cite{Carrillo1996,Carvalho2014_HS_elasticA,Carvalho2014_HS_elasticB}. In
this case $\alpha_2=\Delta \rho \Omega^2 R^5/\mathcal{B}$, where
$\Omega$ is the rotation rate and $\Delta \rho$ is the
density difference between the interior and exterior fluids. In both
cases the wrinkling arises from a competition between the pressure
difference favoring buckling and an opposing force generating
wrinkling with length scale $\lambda$. 

Our finding relies on a remarkable feature of weakly nonlinear theory describing periodic perturbations of the circle solution of Eqs.~(\ref{eq:phieq}) in powers of the perturbation amplitude $\epsilon$~\cite{foster_pressure-driven_2022}: for the case $n=2$, the solution is independent of the force-strength parameter $\alpha_2$ at every order.  Upon expansion of $\phi, T, \textrm{ and }P$, we solve the resulting linear problem at each order in $\epsilon$ and find periodic corrections defined by wave number $m$ and its harmonics to the circle solution:
\begin{align} 
\phi(s) & = s + \epsilon \sin(ms) + \epsilon^2 \frac{\sin(2ms)}{8m} + \mathcal{O}(\epsilon^3)\,.\label{pertphi}
\end{align}
We compute the correction terms in the expansion $(T,P)=(T_0,P_0)+\epsilon^2 (T_2,P_2)+O(\epsilon^4$)   by imposing the solvability condition at each order. Due to the rotational symmetry, only the even orders are non-zero.  The dependence of $P$ and $T$ on $\alpha_2$ is linear at every order: 
\begin{align}
T_{0}= & \frac{1}{2}(1-\alpha_2)-P_{0}\,,\\
P_{2}= & \frac{2m^{4}-9m^{2}+3}{8\left(m^{2}-1\right)^{2}}\alpha_2+\frac{3\left(m^{2}-1\right)}{8}\,,\label{pertP}\\
T_{2}= & \frac{3}{8\left(m^{2}-1\right)}\alpha_2+\frac{3\left(m^{2}+1\right)}{8}\,.\label{pertT}
\end{align}
Higher order expressions for $\phi_j(s)$ and $(T_j,P_j)$ supporting these observations can be found in the Supplementary Material. 

The absence of any $\alpha_2$ dependence in the profile (\ref{pertphi}) has deep physical implications: wrinkle profiles are {\it universal}, i.e., {\it identical} wrinkles can be observed on rings with substrates of different strength, or even for the free ring $\alpha_2=0$, for appropriate values of the pressure $P$ and tension $T$. The transformation $(T,P)\to(T,P)$ is linear in $\alpha_2$ and hence is equivalent to a one-to-one relation between the pressure $P$ and the substrate force measured by $\alpha_2$. Moreover, since the $\alpha_2=0$ problem has closed-form solutions for $\phi(s)$, so does the problem for any $\alpha_2>0$. In the following we demonstrate this fact, and determine the transformation $(T,P)\to(T,P)$ that maps a given wrinkle profile at $\alpha_2>0$ into the same profile for the free ring ($\alpha_2=0$).

\begin{figure}[h]
    \centering
    \includegraphics[width=0.95\linewidth]{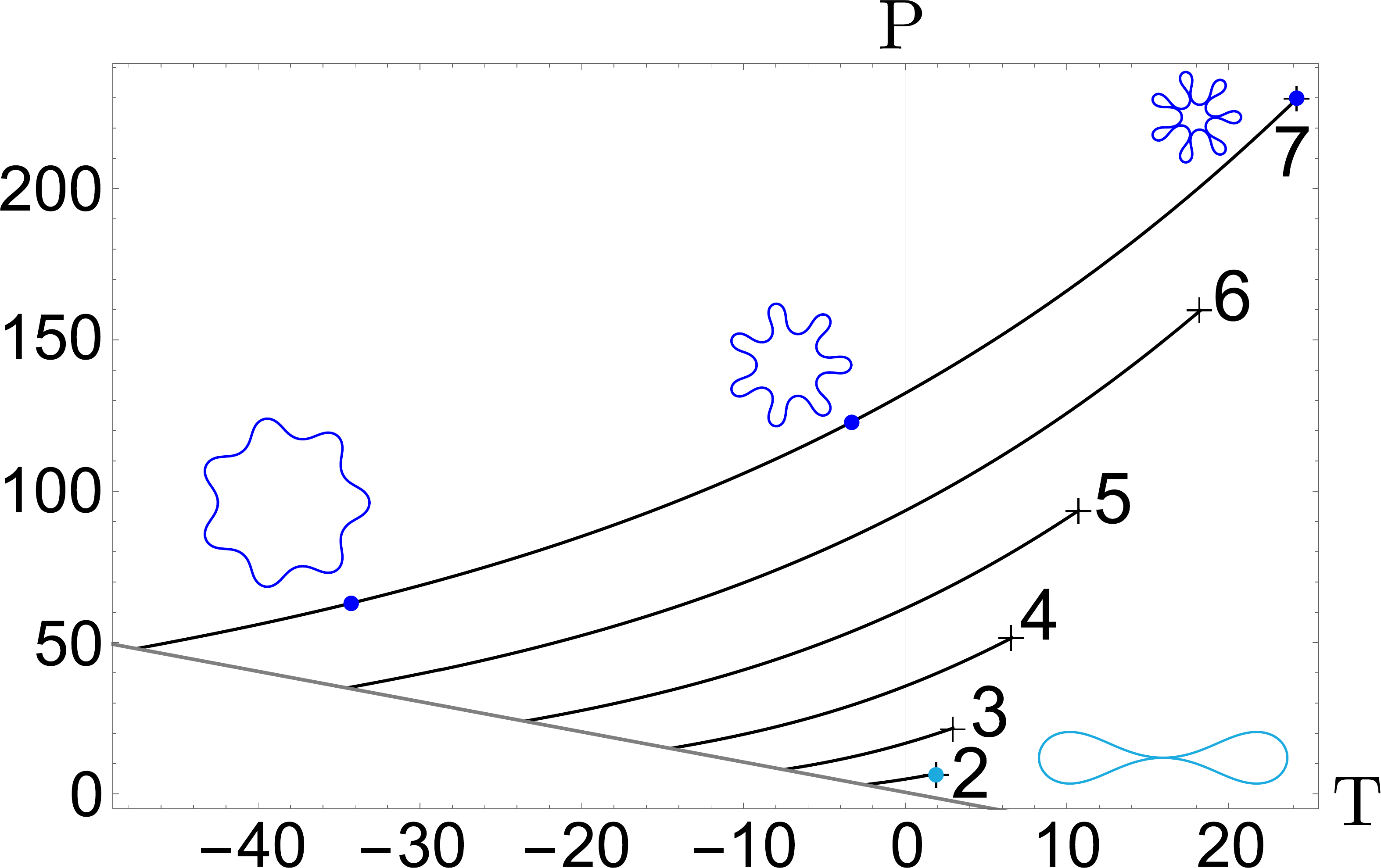}
        \caption{Free-ring buckling for $F=0$. Solutions with wave numbers $2\leq m\leq 7$ (black)
      bifurcate from the circle solution (gray) as $P$ increases starting with $m=2$.  Sample solution profiles for $m=2$ (light blue) and $m=7$ (dark blue) are shown.  Crosses represent points of self-contact.}
    \label{fig:TP0}
\end{figure}

The free inextensible elastic ring problem described by \eqref{eq:phieq} with $F=0$ was studied in detail in
\cite{tadjbakhsh_equilibrium_1967}, and is completely integrable
\cite{Arreaga_shapes,vassilev_cylindrical_2008}. Closed-form
analytical solutions are known and allow the construction of branches of highly nonlinear wrinkle solutions up to the point of self-contact as shown in the ($T,P$) plane in Fig.~\ref{fig:TP0}.  
The wave numbers $m$ come in in the order $m=2,3,\dots$ as $P$ increases above zero, a consequence of the absence of an intrinsic length scale. 

In order to establish the connection between the cases $F=0$ and $F(r)=\alpha_2 r^2$, we consider the equation for $F=0$ with an arbitrary scaling, defining the curvature $Q$ as a function of an arclength $t$ and the tension and pressure parameters $\mu,\sigma$: 
    \begin{equation}
        \frac{d^2 Q(t)}{dt^2} + \frac{1}{2}Q^3(t) - \frac{\mu}{2}Q(t) - \frac{\sigma}{2} = 0\,.
        \label{eq:purebuckle}
    \end{equation}
This equation has the integral
    \begin{equation}
        \left(\frac{dQ}{dt}\right)^2 = 2E - \frac{1}{4}Q^4+\frac{\mu}{2}Q^2 + \sigma Q\,, \label{eq:quartic}
    \end{equation}
where $E$ is the constant of integration.  We also note a key geometric identity satisfied by the corresponding radius $\rho\equiv \sqrt{X^2+Y^2}$,
    \begin{equation}
        \rho^2(t) - \frac{8E+\mu^2}{\sigma^2} - \frac{4Q(t)}{\sigma} = 0\,, \label{eq:Rident}
    \end{equation}
identified in \cite{Arreaga_shapes}.  Equation (\ref{eq:purebuckle}) has exact solutions given by \cite{vassilev_cylindrical_2008}
    \begin{equation}
        Q(t) = \frac{(A\beta+B\alpha)-(A\beta-B\alpha) \textrm{cn} (u t,k)}{(A+B)-(A-B) \textrm{cn} (u t,k)}\,, \label{eq:analytical}
    \end{equation}
where $A,B,u,k$ are functions of the four roots $\alpha < \beta \in \mathbb{R},\gamma=\bar{\delta} \in \mathbb{C}$ of the quartic polynomial on the right side of Eq.~(\ref{eq:quartic}), and $\textrm{cn}(ut,k)$ is the elliptic cosine function with modulus $\sqrt{k}$ (explicit expressions are given in Supplementary Material).  Other solutions exist, but are unphysical owing to self-intersection.  
    \begin{figure*}
        \centering
        \includegraphics[width=0.9\linewidth]{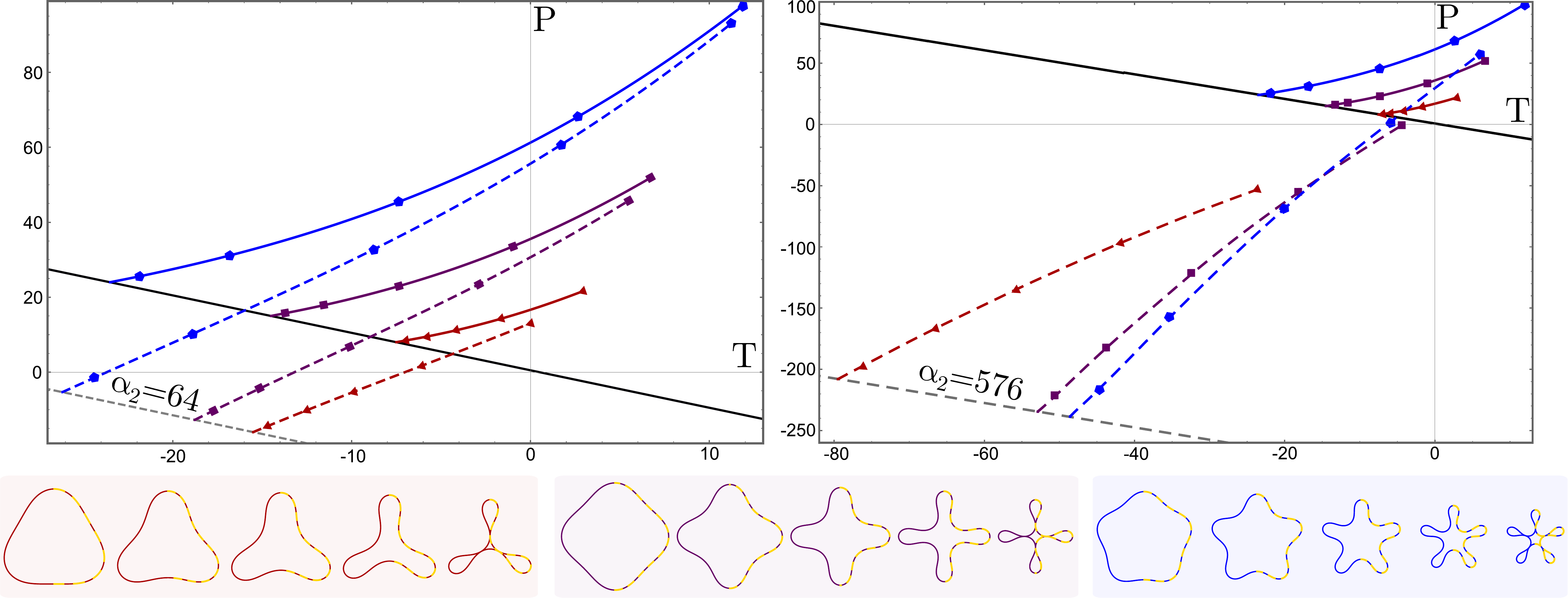}
        \caption{Top: Numerical continuation for two values of
          $\alpha_2$ showing solution branches in the $(T,P)$ plane
          with wave numbers $m=3$ (red triangles), $m=4$ (purple squares) and $m=5$ (blue pentagons). Solid lines: $\alpha_2=0$ (see figure \ref{fig:TP0}); dashed lines $\alpha_2=64$ (left) and $\alpha_2=576$ (right).  Colored markers on the dashed lines map to the corresponding markers on the solid lines.
          Bottom: Color-coded solution profiles at points indicated in the top panels. The solid profiles show the analytical solution while the superposed orange dashed profiles are from numerical continuation (right half of each profile). The solutions agree to within numerical accuracy. In each case, the final profile corresponds to self-contact.}
        \label{fig:mapping}
    \end{figure*}

Finding an exact physical solution to Eq.~(\ref{eq:purebuckle}) then reduces to finding combinations of the three parameters $\mu$, $\sigma$ and $E$ which yield closed, non-self-intersecting curves when employed in Eq.~(\ref{eq:analytical}).
Moreover, adding an appropriate multiple of (\ref{eq:Rident}) to Eq.~(\ref{eq:purebuckle}) and rescaling $t=sR$, $Q=\kappa/R$, $\rho=rR$, we obtain
\begin{align}
     & \frac{d^2\kappa(s)}{ds^2} + \frac{1}{2}\kappa^3(s) -     R^2\left(\frac{\mu}{2}-\alpha_2\frac{2}{\sigma R^5}\right)\kappa(s) \nonumber \\
     & - R^3\left(\frac{\sigma}{2}-\alpha_2\frac{8E+\mu^2}{2\sigma^2 R^5}\right) - \frac{1}{2}\alpha_2 r^2(s)  = 0 \,. \label{eq:alpha2rescale}
\end{align}
There is thus a one-to-one correspondence between Eqs.~\eqref{eq:phieq} and~\eqref{eq:purebuckle} under the mapping:
\begin{subequations}
    \begin{align}
    T & = \frac{\mu R^2}{2} - \frac{2\alpha_2}{\sigma R^3}\,, \label{eq:Tmap} \\
    P & = \frac{\sigma R^3}{2} -  \frac{8E+\mu^2}{2\sigma^2 R^2}\alpha_2 \,, \label{eq:Pmap} \\
     r^2(s) & = \frac{8E+\mu^2}{\sigma^2R^2}+\frac{4\kappa(s)}{\sigma R^3}\,. \label{eq:r}
    \end{align}
    \label{eq:map}
\end{subequations}
Consequently the closed-form analytical solutions of Eq.~(\ref{eq:purebuckle}) also apply to Eqs.~(\ref{eq:phieq}) with $n=2$ and hence describe the wrinkling of rings subject to any force of the form $F \propto r^2$, cf.~\cite{djondjorov_ICGIQ}.
  
For comparison, we extend the $\alpha_2=0$ results in Fig.~\ref{fig:TP0} to nonzero values of $\alpha_2$ using numerical continuation of~(\ref{eq:phieq}) in AUTO \cite{doedel08auto-07p} to show that these correspond to the analytical result (\ref{eq:analytical}) at appropriate locations in the $(T,P)$ plane.

Solving (\ref{eq:map}) using numerically generated values of $x,y,\kappa$ at known $(T,P)$ yields the values of $\mu,\sigma,E$ needed to construct the corresponding analytical solution (\ref{eq:analytical}).  Once we have these solutions, we can use the mapping in Eqs.~(\ref{eq:Tmap},~\ref{eq:Pmap}) to compare the parameter-space location with that of the free-ring problem or to map the free-ring solutions to the corresponding location in parameter space for the substrate-supported ring problem with nonzero $\alpha_2$.  The results for $n=2$ and two values of $\alpha_2$ are shown in Fig.~\ref{fig:mapping} and demonstrate perfect agreement between the numerical continuation results and the closed-form solution of the free-ring problem at the corresponding points in the $(T,P)$ plane. A quantitative comparison (see Supplementary Material) confirms this geometric {\it universality}.

\begin{figure}[h!]
    \centering
    \includegraphics[width=0.65\linewidth]{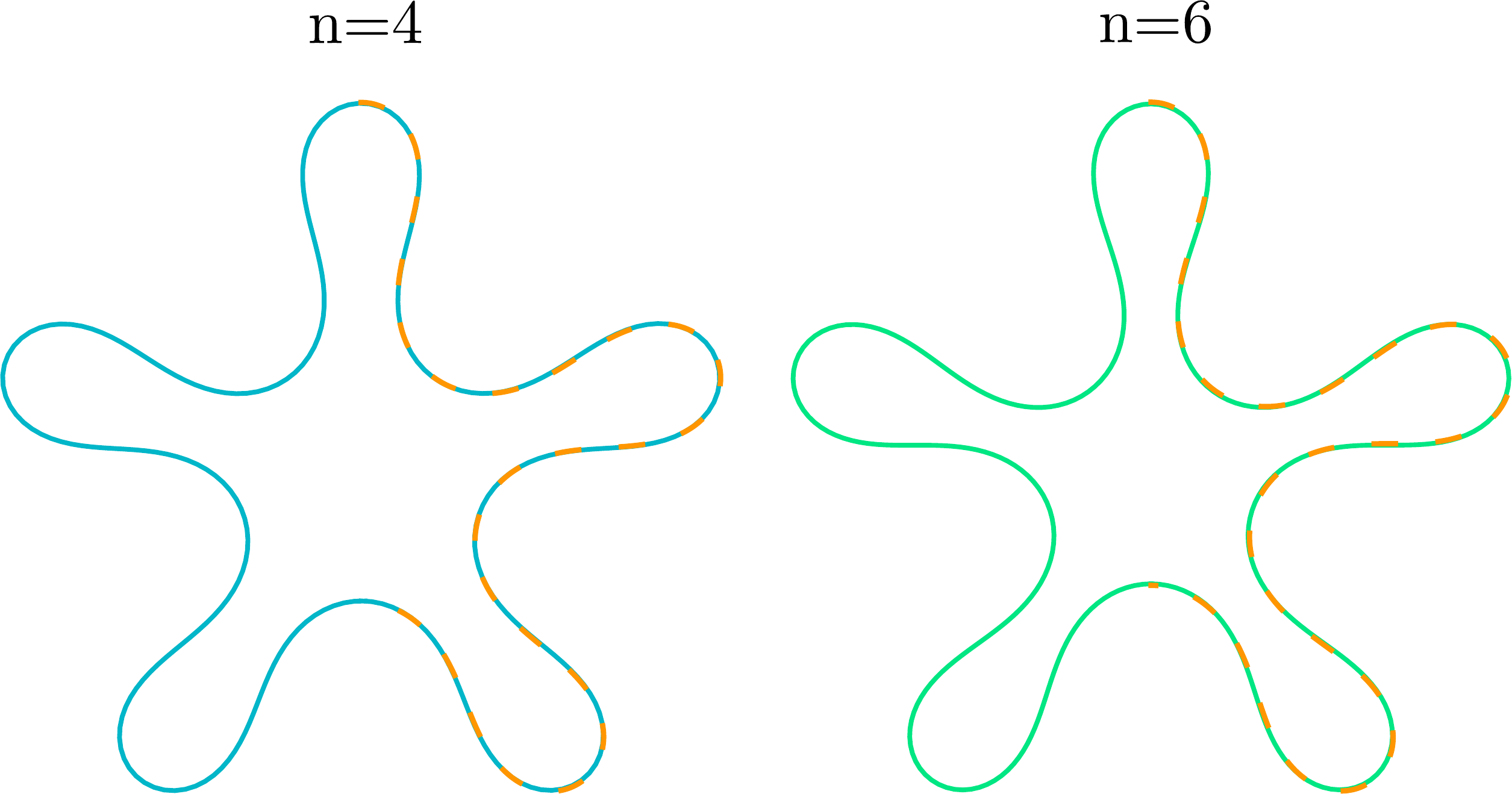}
    \caption{Analytical solutions ($n=4$ blue, $n=6$ green) overlaid with numerical solutions (orange) for $\alpha_{4,6}=500$ and $R=1$.}
    \label{fig:n4n6sol}
\end{figure}

Remarkably, this universality extends beyond $F\propto r^2$: an analogous procedure, involving the addititon of powers of the identity (\ref{eq:Rident}) to Eq.~(\ref{eq:purebuckle}), can be used to map the free-ring solutions onto a broader family of substrate forces including those for which $n=4,6$. A simple translation and rescaling of the curvature (see Supplementary Material) then shows that the free-ring solutions may be used to construct new analytical solutions for both $n=4$ and $n=6$. Figure~\ref{fig:n4n6sol} shows overlays of the resulting analytical and numerical solutions for these values of $n$.

\begin{figure}
    \centering
    \includegraphics[width=0.95\linewidth]{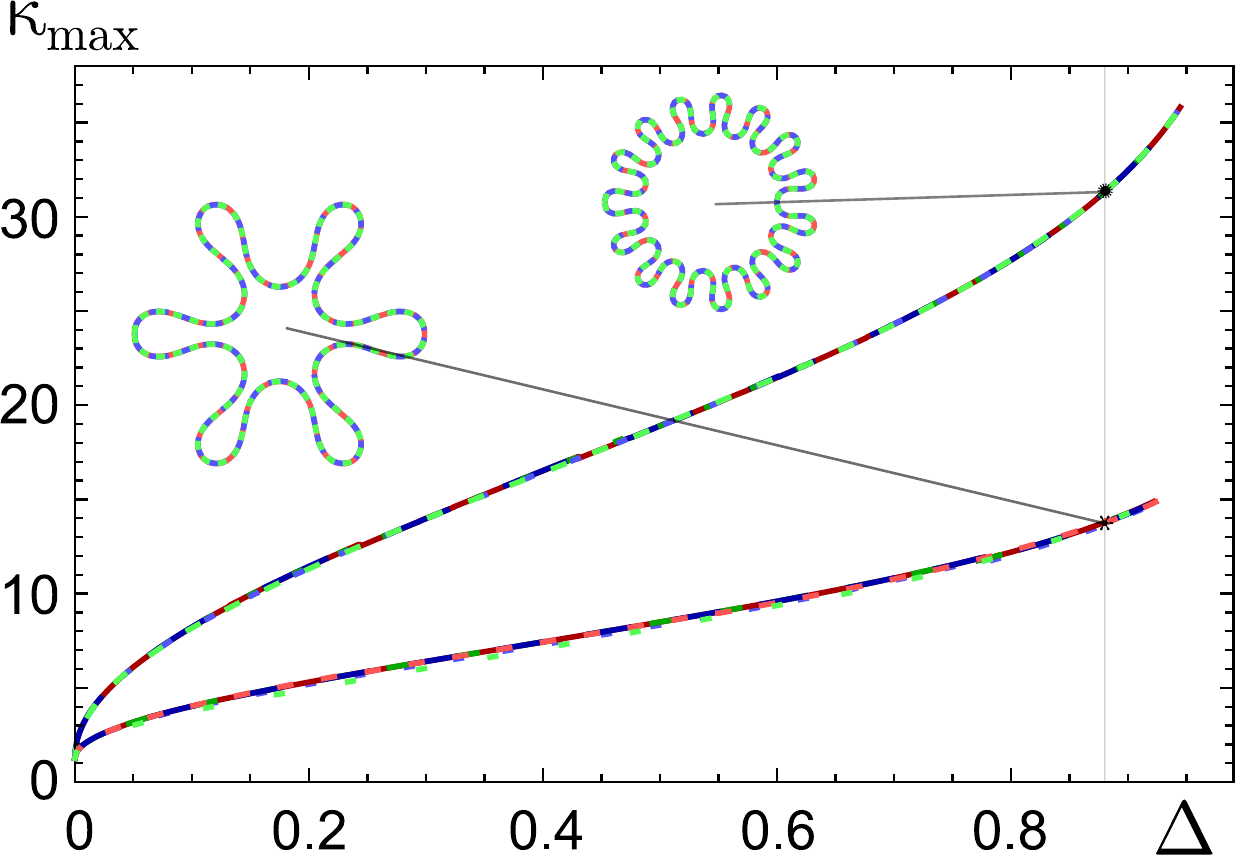}
    \caption{The compression $\Delta \equiv 1-A/\pi$ for area $A$ plotted against maximal curvature $\kappa_{max}$ for $m=6$ and $m=15$.  Solutions for $n=2$ (dark red: $\alpha_2=10$ and light red: $\alpha_2=1000$), $n=4$ (dark blue: $\alpha_4=10$ and light blue: $\alpha_4=1000$), and $n=6$ (dark green: $\alpha_6=10$ and light green: $\alpha_6=1000$). Solutions are shown at $A=0.4$ ($\Delta \approx 0.87$).}
    \label{fig:r2r4r6sol}
\end{figure}

The presence of an additive constant in the curvature results in spatial solutions that are no longer exactly identical, although in the limit of large $R$ or $m$ or small $\alpha_n$, the additive term is heavily suppressed and the solutions approach a universal profile.  In Fig.~\ref{fig:r2r4r6sol} we show how the geometrical features for $n=2,4,6$ compare across a wide range of $\alpha_n$ and $m$.  
The above construction also suggests a straightforward extension to substrate forces of the form $F\sim\sum\alpha_n r^n$, albeit with a more complicated $(T,P)\to(T,P)$ mapping, allowing analytical solution of the wrinkle problem with more complex (and more realistic) substrate forces.

\begin{figure}
    \centering
    \includegraphics[width=0.75\linewidth]{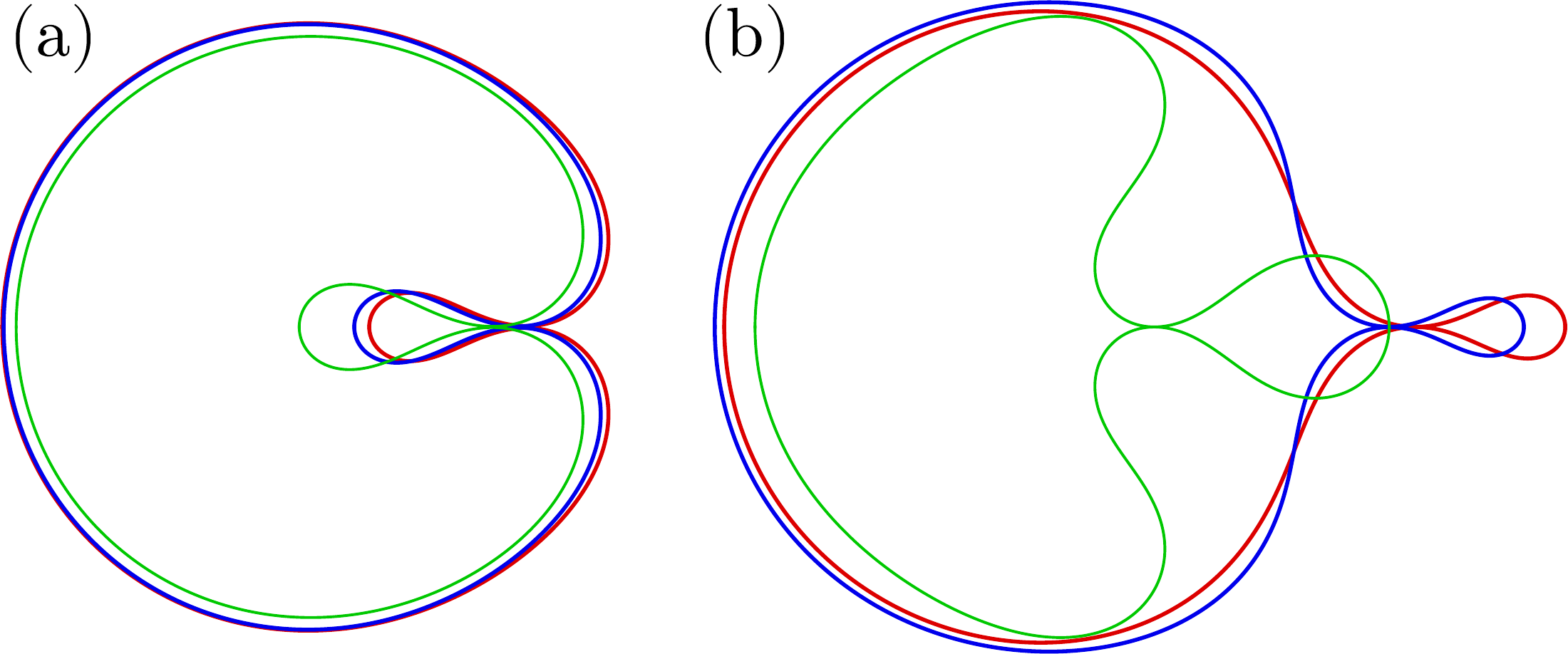}
    \caption{Secondary fold states for $\alpha_n=576$ at the point of self-contact bifurcating from the first primary branch in each case: $n=2$ (red, bifurcates from $m=5$), $n=4$ (blue, bifurcates from $m=6$), and $n=6$ (green, bifurcates from $m=7$). (a) Fold states with intrusion. (b) Fold states with extrusion. The profiles are strongly dependent on the exponent $n$.
    \label{fig:folds}
    }
\end{figure}

Although the wrinkle profiles are the same, the $(T,P)$ mapping modifies the physical response of the system under study as the pressure load increases. In the free-ring problem ($F=0$) the wave numbers $m$ set in monotonically for $P>0$ as $P$ increases so the first mode to bifurcate from the circle branch is the $m=2$ (buckling) mode. As a consequence of the absence of an intrinsic length scale, none of the primary branches ($m=2,3,\dots$) undergoes any secondary bifurcations right up to self-contact.  However, when this length scale is present ($\alpha_n>0$) wrinkle branches may set in in a different order, and the first mode to bifurcate as $P$ increases may have $m>2$ (wrinkle mode) and set in at negative $P$.  Moreover, as $\alpha_n$ increases, secondary
bifurcations move down along the wrinkle branches, eventually passing
the point of self-contact. Thus, for large enough $\alpha_n$, secondary
bifurcations take place prior to self-contact, and these generate
{\it folds} if the resulting secondary branch does not connect to another wrinkle branch or {\it mixed modes} if it does. The mixed modes are characterized by the simultaneous presence of two distinct wave numbers $m_1$, $m_2$ whenever they connect primary branches with wave numbers $m_1$ and $m_2$ \cite{foster_pressure-driven_2022}. When $n=4,6$, similar structures are
observed. Figure~\ref{fig:folds} shows examples of fold states with
intrusion and extrusion, but these are no longer universal and have no counterpart in the free-ring problem with $F=0$. Mixed states are also present when $n=4,6$ but their behavior is complicated by the presence of tertiary bifurcations (not shown). Note that despite the mapping of the wrinkle solutions of (\ref{eq:phieq}) with $\alpha_n>0$ onto the free-ring problem, the presence of folds (and mixed modes) does require substrate support.

We have shown that equations of the form (\ref{eq:phieq}) possess {\it identical} closed-form solutions when $n=0,2$, albeit at different locations in parameter space, and {\it near-identical} closed-form solutions when $n=4,6$. This is despite the presence of an intrinsic length scale when $\alpha_n > 0$. This remarkable result for $n=0,2$ is consistent with the perturbation theory result that $\phi$ is independent of $\alpha_2$ to all orders (see Supplementary Material), while $T,P$ do depend on $\alpha_2$ but only linearly (cf.~Eqs.~(\ref{pertP}) and (\ref{pertT})). These facts suggest that we may differentiate Eq.~(\ref{eq:phieq}) with respect to $\alpha_2$, yielding
\begin{equation}
T_{\alpha_2} \kappa + P_{\alpha_2} + \frac{1}{2} r^2 = 0\,,
\end{equation}
where $T_{\alpha_2},P_{\alpha_2}$ are {\it constants}, an equation that is equivalent to (\ref{eq:r}). Thus the mapping of (\ref{eq:phieq}) onto the free-ring problem applies to the primary wrinkle solutions for which $\phi$ is independent of $\alpha_2$ but not to secondary states where this condition no longer holds. The supported ring problem (\ref{eq:phieq}) is therefore integrable in this limited sense.

This work was supported in part by the National Science Foundation under Grant DMS-1908891 (BF, NV and EK). The work of NV was funded by the National Agency for Research and Development (ANID) through the Scholarship Program: Becas de Postdoctorado en el Extranjero, Becas Chile 2018 No.~74190030. LG was funded by Grant Conicyt Fondecyt Regular 1221103.

\bibliographystyle{apsrev}

\providecommand{\noopsort}[1]{}\providecommand{\singleletter}[1]{#1}%

\end{document}


\title{Supplementary material}
\author{Benjamin Foster, Nicol\'{a}s Verschueren, Edgar Knobloch, and Leonardo Gordillo}
\date{\today}
\maketitle

\section*{}

We consider periodic solutions to the equations
\begin{align}
       & \partial_s^2 \kappa + \frac{1}{2}\kappa^3 - T \kappa - P - \frac{1}{2} F(r) = 0, \nonumber \\
       &\kappa(s)\equiv \partial_s \phi, \quad \partial_s{\bf r} = (\cos \phi, \sin \phi), \quad  F_n(r)=\alpha_n r^n.
       \label{eq:suppeq}
\end{align}

\section*{Exact solutions: $\alpha_n=0$}
Following \cite{vassilev_cylindrical_2008}, the pure buckling problem is defined as
  \begin{equation}
        \frac{d^2 Q(t)}{dt^2} + \frac{1}{2}Q^3(t) - \frac{\mu}{2}Q(t) - \frac{\sigma}{2} = 0
        \label{eq:purebuckle}
    \end{equation}
and the exact solution to this problem is given by
    \begin{equation}
        Q(t) = \frac{(A\beta+B\alpha)-(A\beta-B\alpha) \textrm{cn} (u t,k)}{(A+B)-(A-B) \textrm{cn} (u t,k)}\,, \label{eq:analytical}
    \end{equation}
where    
    \begin{equation}
      A =\sqrt{4\eta^2+(3\alpha+\beta)^2},\qquad  B =\sqrt{4\eta^2+(\alpha+3\beta)^2},\qquad u =\sqrt{AB}/4, \qquad \eta = \frac{\gamma-\delta}{2 i}\,,
    \end{equation}
    and
    \begin{equation}
        k  = \frac{1}{\sqrt{2}}\sqrt{1 - \frac{4\eta^2+(3\alpha+\beta)(\alpha+3\beta)}{[4\eta^2+(3\alpha+\beta)(\alpha+3\beta)]^2+16\eta^2(\beta-\alpha)^2}}.
    \end{equation}
Here $\alpha,\beta,\gamma,\delta$ are the roots of the quartic polynomial (6) given by:
    \begin{align}
        \alpha = & \small{-\sqrt{\frac{\omega}{2}}-\sqrt{\mu-\sigma \sqrt{\frac{2}{\omega}}-\frac{\omega}{2}}} \\
       \beta =  & -\sqrt{\frac{\omega}{2}}+\sqrt{\mu-\sigma \sqrt{\frac{2}{\omega}}-\frac{\omega}{2}} \\
       \gamma =  & \sqrt{\frac{\omega}{2}}+\sqrt{\mu+\sigma \sqrt{\frac{2}{\omega}}-\frac{\omega}{2}} \\
       \delta = & \sqrt{\frac{\omega}{2}}-\sqrt{\mu+\sigma \sqrt{\frac{2}{\omega}}-\frac{\omega}{2}}\,, 
    \end{align}
where $\alpha,\beta \in \mathbb{R}$ and $\gamma = \bar{\delta} \in \mathbb{C}$ are complex, and $\omega$ is a function of the parameters $\mu,\sigma,E$,
\begin{equation}
    \omega = \frac{\left[ \mu + \left(3(3^2\sigma^2+\sqrt{\chi})-\mu\left(\mu^2+2^3 3^2 E\right) \right)^{1/3}\right]^2-2^3 3 E}{3\left(3(3^2\sigma^2 + \sqrt{\chi})-\mu(\mu^2+2^3 3^2 E) \right)^{1/3}}\,,
\end{equation}
and
\begin{equation}
    \chi = 3\left(2^3 E\left[(\mu^2+8E)^2-3^2 2 \mu \sigma^2\right]-\sigma^2(2\mu^3-3^3\sigma^2)\right).
\end{equation}
Other forms of the solution exist, but these necessarily self-intersect.

Either sign of $\sigma$ is valid, as the sign of $\sigma$ may be changed in eq. (\ref{eq:purebuckle}) by changing the sign of $\kappa$. The choice that $\alpha,\beta \in \mathbb{R}$ and $\gamma,\delta \in \mathbb{C}$ requires that $\sigma<0$ and therefore the association $\sigma \propto -P$ such that a positive net external pressure load $P$ leads to buckling solutions.  If, instead, $\sigma>0$ such that $\alpha,\beta \in \mathbb{C}$ and $\gamma, \delta \in \mathbb{R}$, the solution remains physically valid provided the pressure is redefined such that $\sigma \propto P$.  Therefore, the solution applies in both cases assuming the roots are appropriately relabeled and the pressure redefined.

\section*{Rescaling: $r^2$}
We compare our problem (\ref{eq:suppeq}) with $\alpha_2\ne0$ to the buckling problem $\alpha_0=0$ with an extra term which vanishes in view of the geometric identity (8) in the main text.
To do this we write the pure buckling problem (6) in the form
\begin{equation}
    \frac{d^2Q(t)}{dt^2} + \frac{1}{2}Q^3(t) - \frac{\mu}{2}Q(t) - \frac{\sigma}{2} - \frac{1}{2}\frac{\alpha_2}{R^5}\left(\rho^2(t) - \frac{8E+\mu^2}{\sigma^2} - \frac{4Q(t)}{\sigma}\right) = 0 \,. 
\end{equation}
Rearranging, we obtain:
\begin{equation}
   \frac{d^2Q(t)}{dt^2} + \frac{1}{2}Q^3(t) - \left(\frac{\mu}{2}-\frac{2\alpha_2}{\sigma R^5}\right)Q(t) - \left(\frac{\sigma}{2}-\frac{8E+\mu^2}{2\sigma^2 R^5}\alpha_2\right) - \frac{1}{2} \frac{\alpha_2}{R^5} \rho^2(t) = 0 \,. \label{eq:modproblem}
\end{equation}
The problem is then rescaled to map it onto the variables used for $n=2$. For this purpose we assume that $t=sR$, where $R$ is a scaling factor that maps the domain length to $2\pi$ in the variable $s$, so that
    \begin{equation}
        Q = \frac{\kappa}{R}, \quad Q^3 = \frac{\kappa^3}{R^3}, \quad \frac{d^2Q}{dt^2} = \frac{1}{R^3}\frac{d^2 \kappa}{ds^2}, \quad  \rho=Rr\,. \label{eq:rescale}
    \end{equation}
It follows that Eq.~(\ref{eq:modproblem}) then takes the form    
      \begin{equation}
          \frac{d^2\kappa(s)}{ds^2} + \frac{1}{2}\kappa^3(s) -  R^2\left(\frac{\mu}{2}-\frac{2\alpha_2}{\sigma R^5}\right)\kappa(s) - R^3\left(\frac{\sigma}{2}-\frac{8E+\mu^2}{2\sigma^2 R^5}\alpha_2\right) - \frac{1}{2}\alpha_2 r^2(s)  = 0\,,
          \end{equation}
allowing a direct comparison with (\ref{eq:suppeq}). Thus
\begin{align}
    T & = \frac{\mu R^2}{2} - \frac{2\alpha_2}{\sigma R^3}\,, \\
    P & = \frac{\sigma R^3}{2} - \frac{8E+\mu^2}{2\sigma^2 R^2}\alpha_2\,.
\end{align}
The result demonstrates that the solutions of (\ref{eq:suppeq}) with $\alpha_2>0$ and those with $\alpha_0=0$ are in fact identical, provided the parameters are adjusted accordingly.

\section*{Rescaling: $r^4$}
We can repeat the above procedure, adding different terms which are identically zero to the pure buckling equation.  For example, squaring the geometric identity (8) and adding the result to (\ref{eq:purebuckle}) results in
\begin{equation}
       \frac{d^2Q(t)}{dt^2} + \frac{1}{2}Q^3(t) - \frac{\mu}{2}Q(t) - \frac{\sigma}{2} - \frac{1}{2}\frac{\alpha_4}{R^7}\left(\rho^4(t) - \left(A + BQ(t)\right)^2\right) = 0 \,,
\end{equation}
where $A \equiv \frac{8E+\mu^2}{\sigma^2}$ and $B \equiv \frac{4}{\sigma}$. This equation has a term quadratic in $Q$ which does not match the form of (\ref{eq:suppeq}), but by writing $Q=\tilde{Q}+C$ we can eliminate this term with the choice $C=-\frac{ \alpha_4 B^2}{3 R^7}$.  The same rescaling as in Eq.~(\ref{eq:rescale}) can now be performed to obtain
\begin{equation}
     \kappa_{ss} + \frac{1}{2}\kappa^3 -   \left(\frac{\mu  R^2}{2} - \frac{\alpha_4 A B}{R^5} + \frac{\alpha_4^2 B^4}{6 R^{12}} \right) \kappa - \left(\frac{\sigma R^3}{2} - \frac{\alpha_4}{2R^4} \left(A^2 +\frac{\mu B^2}{3}\right) + \frac{A \alpha_4^2 B^3}{3 R^{11}} -\frac{\alpha_4^3 B^6}{27 R^{18}}\right) - \frac{1}{2} \alpha_4 r^4 = 0\,.
\end{equation}
The mapping to the original problem is obtained by identifying $T,P$ with the quantities in the first and second parentheses, respectively, and incorporating the definitions for $A,B$:
\begin{align}
    T & = \frac{\mu  R^2}{2} - \frac{4 \alpha_4 \left(\mu ^2+8 E\right)}{R^5 \sigma ^3} + \frac{128 \alpha_4^2}{3 R^{12} \sigma ^4}\,,  \\
    P & = \frac{ \sigma R^3 }{2} - \frac{\alpha_4}{R^4\sigma^4} \left(\frac{\left(\mu ^2+8 E\right)^2}{2} + \frac{8 \mu \sigma^2}{3 }\right) + \frac{64 \alpha_4^2 \left(\mu ^2+8 E\right)}{3 R^{11} \sigma ^5}  - \frac{4096 \alpha_4^3}{27 R^{18} \sigma ^6}\,.
\end{align}
This procedure demonstrates that when $n=4$ the wrinkle solutions of (\ref{eq:suppeq}) and the buckling solutions of (\ref{eq:purebuckle}) are still identical, to within an additive constant in the curvature, provided the parameters $(T,P)$ are adjusted accordingly.

\begin{figure}
    \centering
    \includegraphics[width=0.25\linewidth]{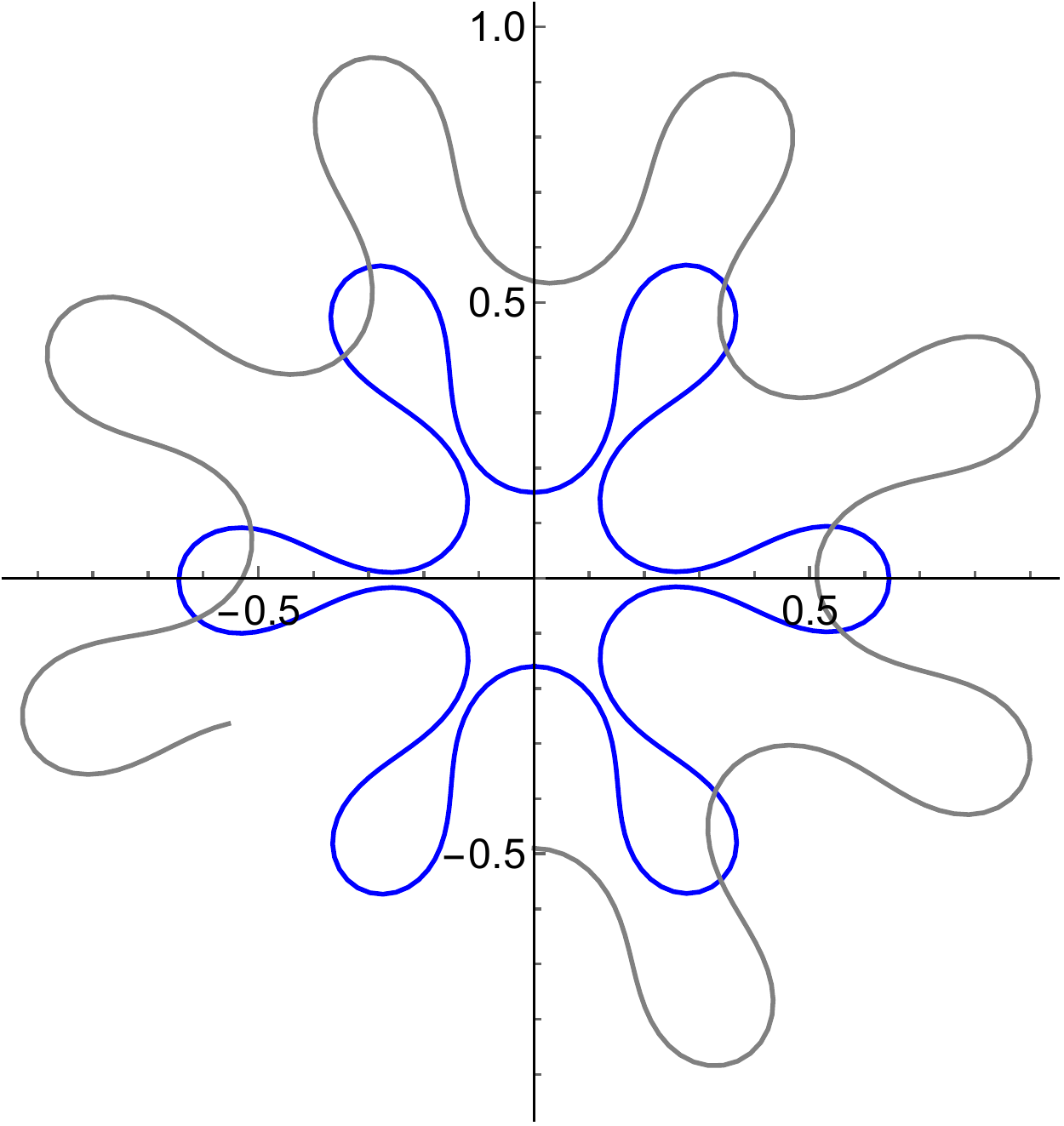}
    \caption{Spatial profile resulting from the analytical solution $\kappa(s)$ solving (\ref{eq:suppeq}) for $\alpha_4=576$ (blue) and that from $Q(t)$ solving (\ref{eq:purebuckle}) (gray) for the same parameters: $\mu=0.52,\sigma=100.45, E=694, R=1.14176$. Owing to the translation term relating the curvatures $\kappa(s)$ and $Q(t)$, only $\kappa$ closes smoothly.}
    \label{fig:r4comp}
\end{figure}

With $Q(t)$  given by (\ref{eq:analytical}) as the solution to the pure-buckling problem ($\ref{eq:purebuckle}$), we define the solution to the $\alpha_4>0$ problem via $\kappa$ where $\kappa = R(Q-C)$.  To maintain a closed physical solution, we once again find parameters for $Q(t;\mu, \sigma, E)$ but this time we require that $\int_0^{2\pi R} \kappa ds = \int_0^{2\pi R} R\left(Q +\frac{16\alpha_4}{3\sigma^2 R^7}\right) ds= 2\pi$.  Profiles of $\kappa$ and the corresponding $Q$ from which it is derived are shown in Fig. \ref{fig:r4comp}. Only $\kappa$ is closed.

\section*{Rescaling: $r^6$}

Proceeding in a similar manner, we now add a cube of the geometric identity (8) to eq. (\ref{eq:purebuckle}), obtaining
\begin{equation}
       \frac{d^2Q(t)}{dt^2} + \frac{1}{2}Q^3(t) - \frac{\mu}{2}Q(t) - \frac{\sigma}{2} - \frac{1}{2}\frac{\alpha_6}{R^7}\left(\rho^6(t) - \left(A + B Q(t)\right)^3\right) = 0 \,,
\end{equation}
where $A,B$ are as before.  The problem may be transformed into the form (\ref{eq:suppeq})
with $n=6$ via the transformation: $Q=\kappa/(\gamma R)+C$, $t=sR$, $\rho=Rr$, where $C=-\frac{\alpha_6 16(8E+\mu^2)}{64 \alpha_6 \sigma + R^9 \sigma^4}$ and $\gamma = 1/\sqrt{1+64\alpha_6/R^9\sigma^4}$.  Inserting the definitions of $A,B$ from above, we obtain:
\begin{align}
& \kappa_{ss} + \frac{1}{2} \kappa^3  - \kappa \frac{(
\frac{1}{2} R^2(\mu ( \sigma^8 R^{18} + 4096 \alpha_6^2\sigma^2 + 
       128 R^9 \alpha_6\sigma^5) - 
    768 \alpha_6 (8E + \mu^2)^2 ( \alpha_6 + 
       \frac{R^9 \sigma^3}{64}))))}{( R^9 \sigma^4 + 
   64 \alpha_6\sigma)^2} \\
  & - 
 \frac{1}{2} \sqrt{1 + \frac{64 \alpha_6}{R^9 \sigma^3}} \left(\sigma R^3
 + \frac{
 16 R^3 \alpha_6 \mu (8 E + \mu^2) }{(
 64 \alpha_6 \sigma + R^9 \sigma^4)} + \frac{
  (8E+ \mu^2)^3  (-4096 R^3 \alpha_6^3 + 
     R^{21} \alpha_6 \sigma^6)}{(64 \alpha_6 \sigma + 
   R^9 \sigma^4)^3}\right) \\
   &   - 
 \frac{1}{2}  \sqrt{1 + \frac{64 \alpha_6}{R^9 \sigma^3}} \alpha_6 r^6 = 0\,.
\end{align}

\par We use the transformations described above for $n=2,4,6$ to obtain exact solutions of (\ref{eq:suppeq}) in terms of the exact solutions (\ref{eq:analytical}) of the pure buckling problem, and compare the results with numerical continuation results in the next section.

\section*{Quantitative comparisons}
\begin{figure}[h!]
    \centering
    \includegraphics[width=0.9\linewidth]{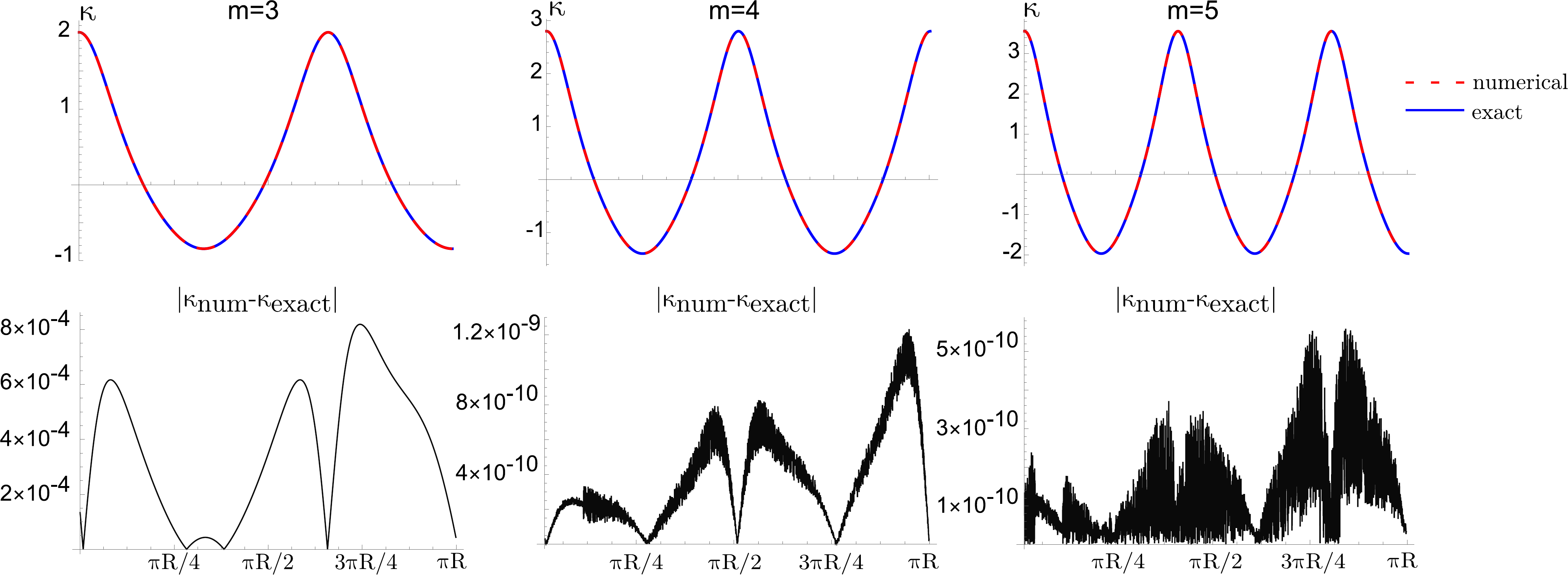}
    \caption{Top: Exact (blue) solution for the curvature $\kappa(s)$ over half the domain for $\alpha_2=576$ and $R=3.41083$ overlaid with the corresponding numerical result (red).  Bottom: absolute difference between the exact and numerically computed $\kappa(s)$.  The error in each case is consistent with the numerical tolerances used in the computation. For the analytical solution the corresponding parameters are as follows: ($m=3$) $\mu=0.50778, \sigma=1.09113, E=0.43376$; ($m=4$) $\mu=1.15369, \sigma=2.61283, E=1.74328$; ($m=5$) $\mu=2.04365, \sigma=4.93013, E=4.76243$.  All three solutions correspond to first self-contact.
    }
    \label{fig:comps}
\end{figure}

We can compare the numerical and analytical results quantitatively for $n=2$ and find excellent agreement between them as demonstrated in Fig.~\ref{fig:comps}.   

\section*{Weakly nonlinear analysis: $n=2$}
It is noted in the main text that the solution constructed from a weakly nonlinear analysis about the circle state is independent of the key parameter $\alpha_2$ at every order computed for $n=2$.  Expressions for $\phi$ to order $\epsilon^6$ are provided below:
\begin{align}
    \phi & = s + \epsilon \sin(ms) + \epsilon^2 \frac{\sin(2ms)}{8m} + \epsilon^3 \frac{(m^2+3)\sin(3ms)}{192 m^2} + \epsilon^4 \frac{\sin(2ms)\left(\cos(2ms)m^2 + 5m^2 + \cos(2ms)-1\right)}{256 m^3} \nonumber \\
    & + \epsilon^5 \frac{\left(m^4+18m^2-3\right)\sin(3ms) + \sin(5ms)\left(\frac{m^4}{5}+2m^2+1\right)}{4096 m^4} \nonumber \\
    & + \epsilon^6 \frac{\sin(2ms)\left(\left(m^4+\frac{10}{3}m^2+1\right)\cos(2ms)^2+\left(6m^4+12m^2-2\right)\cos(2ms)+\frac{109m^4}{4}-\frac{83m^2}{6}+\frac{5}{4}\right)}{8192 m^5} + \mathcal{O}(\epsilon^7).
\end{align}
Further, the dependence of $P$ and $T$ on $\alpha_2$ is linear at every order computed.  We provide these results to order $\epsilon^6$:

\begin{align}
    T & = \frac{\alpha_2}{(m-1) (m+1)}-m^2+\frac{3}{2} \nonumber \\
    & + \left(\frac{3}{8} \left(m^2+1\right)-\frac{3 \alpha_2}{(8 (m-1)) (m+1)}\right) \epsilon^2 \nonumber \\
    & +   \left(\frac{3 \alpha_2 (3 m-1) (3 m+1)}{512 \left(((m-1) (m+1)) m^2\right)}+\frac{3 \left(m^4+22 m^2+1\right)}{512 m^2}\right) \epsilon^4 \nonumber \\
  & +\left(\frac{3 \left(5 m^6+371 m^4+43 m^2-3\right)}{32768 m^4}-\frac{9 \alpha_2 \left(3 m^2+1\right) \left(5 m^2-1\right)}{32768 \left(((m-1) (m+1)) m^4\right)} \right) \epsilon^6 + ... \\ 
    P & = \frac{\alpha_2 \left(m^2-3\right)}{2 ((m-1) (m+1))}+(m-1) (m+1) \nonumber \\
    & + \left(\frac{3}{8} (m-1) (m+1)-\frac{\alpha_2 \left(2 m^4-9 m^2+3\right)}{8 \left((m-1)^2 (m+1)^2\right)} \right) \epsilon^2 \nonumber \\
  & + \left(\frac{\alpha_2 \left(24 m^6-107 m^4+54 m^2-3\right)}{512 \left(\left(m^2 (m-1)^2\right) (m+1)^2\right)}+\frac{3 (m-1) (m+1) \left(15 m^2+1\right)}{512 m^2} \right) \epsilon^4 \nonumber \\
  & + \left(\frac{9 (m-1) (m+1) \left(21 m^2-1\right) \left(3 m^2+1\right)}{32768 m^4}-\frac{\alpha_2 \left(146 m^8-641 m^6+115 m^4+69 m^2-9\right)}{32768 \left(\left(m^4 (m-1)^2\right) (m+1)^2\right)} \right) \epsilon^6 + ... \label{appendix:Pwnla}
\end{align}
We have used computer algebra to extend these results to order $\epsilon^{12}$ and dependence on $\alpha_2$ holds at every order.

\section{Weakly nonlinear analysis $n=4,6$ }
In the cases $n=4$ and $n=6$, $\phi$ is no longer independent of
$\alpha_4$ or $\alpha_6$, respectively.  Similarly, the expressions
for $T,P$ contain nonlinear dependence on $\alpha_4,\alpha_6$.

{\tiny{\begin{align}
  \phi(s)&=s+\sin\! \left(m s\right) \epsilon +\frac{\left(12
                 m^{6}-27 m^{4}+22 \alpha_4 \,m^{2}+18 m^{2}-10
                 \alpha_4-3\right) \sin\! \left(2 m s\right)
                 \epsilon^{2}}{24 m \left(m-1\right) \left(m+1\right)
                 \left(4 m^{4}-5 m^{2}+2
                 \alpha_4+1\right)}+O(\epsilon^3),\\
  & \nonumber \\
T&=\frac{2 \alpha_4}{\left(m-1\right)
         \left(m+1\right)}-m^{2}+\frac{3}{2} \nonumber\\
  &-\frac{\left(\left(84 m^{4}-280
    m^{2}+100\right)\alpha_4^{2}+6 \left(m-1\right)
    \left(m+1\right) \left(25 m^{6}-92 m^{4}+53 m^{2}-10\right)
    \alpha_4-9 \left(2m+1\right) \left(2 m-1\right)
    \left(m^{2}+1\right)\left(m-1\right)^{4}
    \left(m+1\right)^{4}\right) \epsilon^{2}}{24\left(m-1\right)^{3}
    \left(m+1\right)^{3} \left(2\alpha_4+\left(m-1\right)
    \left(m+1\right) \left(2m-1\right) \left(2 m+1\right)\right)}\\
               &+O(\epsilon^4), \nonumber \\
    & \nonumber \\
  P&=\frac{\left(m^{2}-5\right)\alpha_4}{2\left(m-1\right)\left(m+1\right)}+\left(m-1\right)
           \left(m+1\right)\nonumber \\
  &-\frac{\left(4 \left(6 m^{6}-57 m^{4}+100 m^{2}-25\right)
    \alpha_4^{2}+6 \left(m-1\right) \left(m+1\right) \left(8
    m^{8}-81 m^{6}+156 m^{4}-69 m^{2}+10\right) \alpha_4-9
    \left(2 m+1\right) \left(2 m-1\right) \left(m-1\right)^{5}
    \left(m+1\right)^{5}\right) \epsilon^{2}}{24 \left(2
    \left(m-1\right)^{3} \left(m+1\right)^{3} \alpha_4+\left(2
    m-1\right) \left(2 m+1\right) \left(m-1\right)^{4}
    \left(m+1\right)^{4}\right)}\\
  &+O(\epsilon^4).\nonumber
\end{align}}}

\bibliographystyle{apsrev}
\bibliography{exsol}